\documentclass[reprint,aps]{revtex4-1}                  
\usepackage{graphicx}
\usepackage{colortbl}
\usepackage{wasysym}
\usepackage{amssymb, latexsym, textcomp,  textcomp}
\begin{document}

\title{Transport properties of a quasi-1D Wigner Solid on liquid helium confined in a microchannel with periodic potential}


\author {J.-Y. Lin$^\dagger$, A. V. Smorodin, A. O. Badrutdinov, and D. Konstantinov}

 \email{jui-yin.lin@oist.jp}

\affiliation{Quantum Dynamics Unit, Okinawa Institute of Science and Technology (OIST) Graduate University, Tancha 1919-1, Okinawa 904-0412, Japan.\\}


\date{Received: date / Accepted: date}

\begin{abstract}

We present transport measurements in a quasi-1D system of surface electrons on liquid helium confined in a 101-$\mu$m long and 5-$\mu$m wide microchannel where an electrostatic potential with periodicity of $1$-$\mu$m along the channel is introduced. In particular, we investigate the influence of such a potential on the nonlinear transport of quasi-1D Wigner Solid (WS) by varying the amplitude of the periodic potential in a wide range. At zero and small values of amplitude, quasi-1D WS in microchannel shows expected features such as the Bragg-Cherenkov scattering of ripplons and reentrant melting. As the amplitude of potential increases, the above features are strongly suppressed. This behavior suggests loss of the long-range positional order in the electron system, which is reminiscent of the re-entrant melting behaviour due to the lateral confinement of WS in the channel.                   


\end{abstract}
\maketitle

\section{Introduction}

A lattice of interacting charged particles driven against an external periodic potential presents an attractive system to model sliding friction, which is relevant to many diverse fields of science~\cite{Van2013}. The basic model that describes sliding friction between crystalline interfaces, the one-dimensional (1D) Frenkel-Kontorova (FK) model, consists of a chain of particles coupled by a harmonic nearest neighbor interaction and subject to an external spatially periodic potential. The competition between these two interactions determines essential physical features, such as the stick-slip motion when particles are acted upon by an additional adiabatically increasing driving force, the Aubry transition to {\it superlubricity} state, etc. It is understood that these processes are governed by excitation of topological defects in the system and are strongly influenced by incommensurate periodicities of the lattice and potential~\cite{FK-book}. Experimentally, the FK model was studied in diverse physical systems such as solid interfaces~\cite{Die2004,Soc2006} and cold ions in optical lattices~\cite{Bil2015,Bil2016}, and this subject continues attracting a lot of attention. 

Electrons on helium present an ultra-clean system of charged particles on a liquid substrate where the FK model can be potentially realized. Owing to the strong unscreened Coulomb interaction between electrons, the system crystallizes into a two-dimensional Wigner Solid (WS) at relatively low densities of order $10^{13}$~m$^{-2}$ and at temperatures around 1~K. Electrostatic pressure exerted by the localized electrons in the solid phase causes a commensurate deformation of liquid surface, the so called dimple lattice, that couples to WS and significantly alters its transport along the surface under an external driving electric field. In particular, the coherent Bragg-Cherenkov (BC) emission of surface capillary waves, ripplons, with the wave length equal to the electron lattice constant leads to deepening of the dimples and results in saturation of electron current~\cite{Kri1996,Dyk1997}. At high driving fields, the decoupling of electrons from dimples occurs, and WS is in a sliding state characterized by much larger values of current~\cite{Shir1995,Andrei}. The regime of coherent BC emission of ripplons where the lattice of electrons couples to the commensurate deformation of liquid surface has a loose analogy with the two-dimensional (2D) case of FK model for the particular case when the mean distance between particles equals to the spatial period of the substrate potential. Indeed, it shows some typical features predicted by the FK model, such as a depinning transition, hysteresis, etc~\cite{VanossiPRL2008,WangAPL2008}. However, one of the main aspect of FK model, that is the role of incommensurability between electron and dimple lattices, is automatically eliminated by the origin of the dimple lattice itself. 

Recently, it was demonstrated that confining electrons in capillary-condensed microchannel structures facilitates control of the electron system by imposed electrostatic potentials and allow to observe new interesting features associated with the electron transport and phase transitions in the system, such as a clocked electron transport \cite{Brad2011}, discrete transport through a point-contact constriction \cite{Rees2011,Rees2012}, suppressed and re-entrant melting of a quasi-1D electron crystal~\cite{Ikeg2010,Ikeg2012,Rees2013}, stick-slip motion of WS~\cite{Rees2016}, inhomogeneous WS~\cite{BadrPRB2016}, etc. Motivated by these works, and by the possibility to study the FK model in the electron-on-helium system, we designed and fabricated a microchannel device in which a spatially periodic electrostatic potential of varying amplitude could be imposed onto the electrons confined in the channel. Here, we present results of our preliminary studies of the transport of electrons along the microchannel in the presence of the periodic potential. In particular, we report the strong suppression of typical features associated with the crystalline ordering, such as the reentrant melting of WS and nonlinear BC regime of electron transport, by the periodic potential. This results are discussed in terms of loss of the long-range crystalline ordering in the electron system under application of the external potential.        

\section{Sample and methods}

\begin{figure*}
\includegraphics[width=1.0\textwidth]{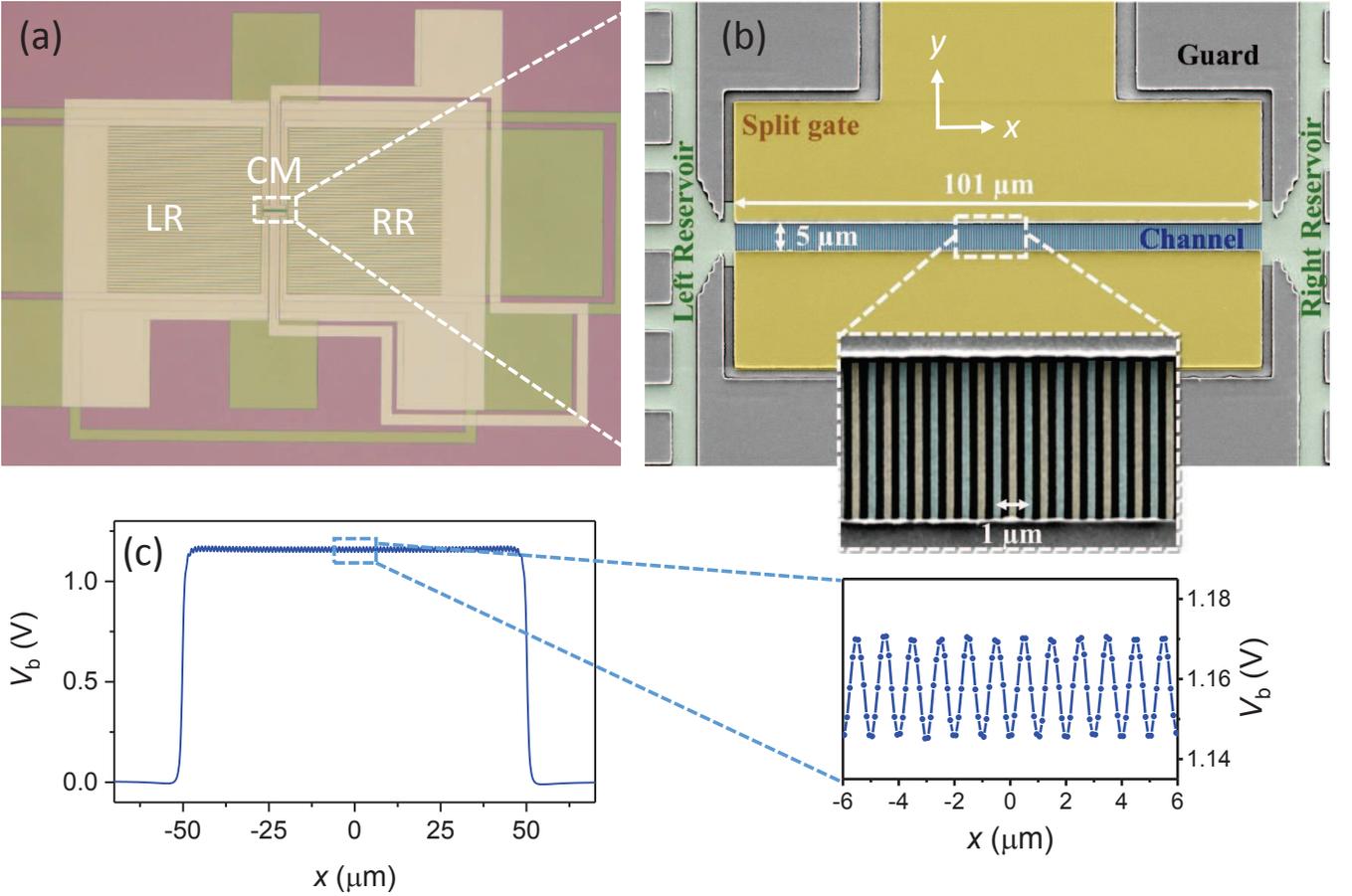}
\caption{(color online)(a)  Microscopic image of the microchannel device used in the experiment. Bottom and top gold layers are shown by dark and light golden colors, respectively. The central microchannel (CM) connecting  the left and right reservoirs, LR and RR respectively, is shown in details in the next panel. (b) False-color scanning electron micrograph of the CM region of the device. The bottom of the channel is formed by two electrically isolated electrodes, Ch1 and Ch2, in the shape of inter-digital capacitor plates. The inset shows magnified image of these plates. (c) Variation of electrostatic potential along the microchannel ($x$-direction) calculated using FEM at the surface of liquid helium in the middle of CM for $V_\textrm{sg}=-0.8$~V, $V_\textrm{ch}=1.5$~V, and $\Delta V_\textrm{ch}=0.5$~V. The inset shows magnified portion of the potential at the surface at the center of CM.}
\label{fig:1}       
\end{figure*}

Our microchannel device was fabricated on a silicon-oxide substrate using the optical and e-beam lithography methods. The device consisted of two large arrays of 5~$\mu$m-wide microchannels, which acted as two electron reservoirs, connected by a central 101~$\mu$m-long and 5~$\mu$m-wide microchannel, see Figs.~\ref{fig:1}(a) and \ref{fig:1}(b). This structure was composed of two patterned gold layers separated by an insulating layer of silicon-nitride having thickness of 550~nm. The bottom gold layer (dark golden color in Fig.~\ref{fig:1}(a)) consisted of three electrodes which defined the bottoms of two reservoirs and the central channel. The top gold layer (light golden color in Fig.~\ref{fig:1}(a)) consisted of two electrodes, the split-gate and guard electrodes. The insulating layer separating two gold layers was removed within the microchannels to form the rectangular-shape groves having height of 550~nm. The microchannels were filled with superfluid $^4$He by capillary action from bulk liquid helium, the level of which was maintained slightly below the device. 

A special feature of our device is the bottom electrode of the central channel, which consisted of two separate parts, "Ch1" and "Ch2", in the shape of an interdigital capacitor with fingers aligned across the channel, see inset of Fig.~\ref{fig:1}(b). Each finger was $250$~nm-wide and adjacent fingers of two electrodes where separated by $250$~nm-wide gaps. Thus, by applying potential difference $\Delta V_\textrm{ch}=|V_\textrm{ch1}-V_\textrm{ch2}|$ between electrodes Ch1 and Ch2, we could create a spatially periodic electrostatic potential along the channel with period of 1~$\mu$m. As an example, Fig.~\ref{fig:1}(c) shows spatial variation of potential at the center of the channel at a distance of $550$~nm above the bottom electrode (that is approximately at the level of liquid helium filling the channel) calculated using the finite element method (FEM) for $\Delta V_\textrm{ch}=0.5$~V. In this case, voltages $V_\textrm{ch1}=V_\textrm{ch}+\Delta V_\textrm{ch}/2$ and $V_\textrm{ch2}=V_\textrm{ch}-\Delta V_\textrm{ch}/2$ were assigned to electrodes Ch1 and Ch2, respectively, to have a common bias of 1.5~V at the bottom of the central microchnnel.  

The surface of liquid helium filling the microchannels was charged with electrons produced by thermal emissions from a tungsten filament placed a few millimeters above the device, while a positive bias was applied to the reservoir's bottom electrodes and the guard electrode was grounded. The transport of electrons through the central microchannel was measured by the standard capacitive-coupling (Sommer-Tanner) method. An ac voltage $V_\textrm{ac}$ at the frequency $f=99.5$~kHz was applied to one of the reservoir electrodes, while both in-phase and quadrature components of the current $I$ induced by electron motion in the other reservoir electrode was measured with a lock-in amplifier. The response of the device was well described by a lumped $RC$-circuit~\cite{Ikeg2015}, where two capacitances ($\sim 1$~pF) between left and right reservoir electrodes and charged surface of liquid are connected in series with resistance $R$ of the electrons in central channel. Note that due to the large size of reservoirs, which consisted of a large number of microchannels connected in parallel, the total resistance of the device is dominated by the resistance of electrons in the central microchannel, which justifies applicability of the above lumped-$RC$ model. In turn, $R$ depends on the density of electrons in the central microchannel, which is determined by the amount of electrons in the reservoirs and the voltages applied to different electrodes of the device.

\section{Experimental results}

\subsection{Phase diagram of electron system without applying periodic potential}

\begin{figure*}
\includegraphics[width=1.0\textwidth]{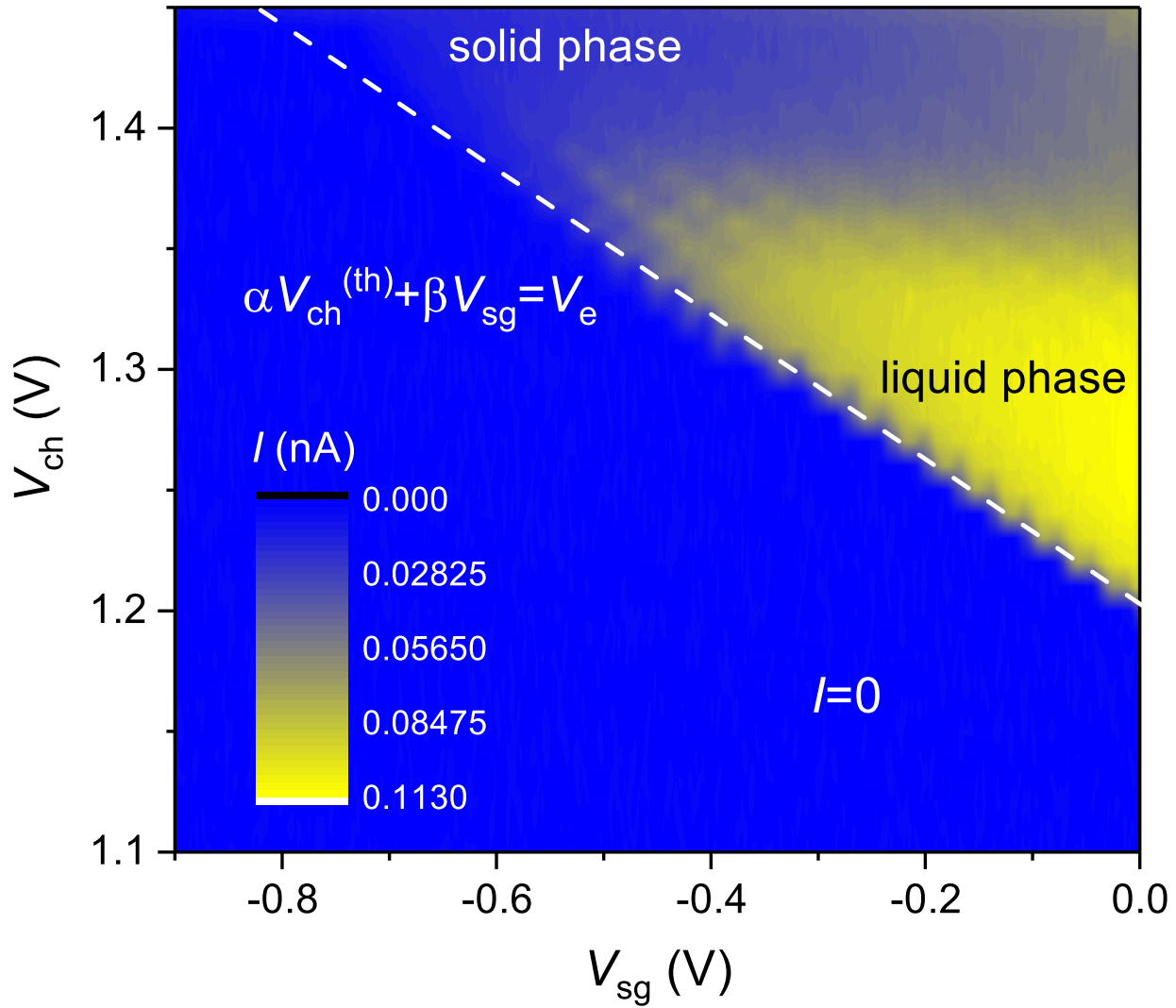}
\caption{(color online) Magnitude of electron current $I$ measured as a function of split-gate electrode potential $V_\textrm{sg}$ and channel potential $V_\textrm{ch}=V_\textrm{ch1}=V_\textrm{ch2}$ at $T=0.86$~K. Dashed (white) line shows potential theshold values for channel opening, as described in the text.}
\label{fig:2}       
\end{figure*}

First, we check performance of the fabricated device by applying the same potential to both electrodes Ch1 and Ch2 of the central microchannel, $V_\textrm{ch}=V_\textrm{ch1}=V_\textrm{ch2}$, and measuring current in the device $I$ while applying the peak-to-peak ac voltage $V_\textrm{ac}=5$~mV to the device. The absolute value of measured $I$ is plotted in Fig.~\ref{fig:2} for various values of $V_\textrm{ch}$ and bias $V_\textrm{sg}$ applied to the split-gate electrode of the central microchannel. To understand this diagram, it is convenient to use a simplified capacitance model to find a relation between the density of electrons in the central micochannel and voltages applied to different electrodes of the device~\cite{Rees2012}. First, we define the total capacitance of the liquid surface in the central microchannel $C_\Sigma=C_\textrm{ch}+C_\textrm{sg}$, where $C_\textrm{ch}$ and $C_\textrm{sg}$ are capacitances between the liquid surface and channel's bottom and split-gate electrodes, respectively.  It is also convenient to introduce the dimensionless coupling constants $\alpha=C_\textrm{ch}/C_\Sigma$ and $\beta=C_\textrm{sg}/C_\Sigma$, which satisfy the obvious relation $\alpha+\beta=1$. Then, the potential at the uncharged liquid surface can be written as $V_\textrm{b}=\alpha V_\textrm{ch}+\beta V_\textrm{sg}$. When the device is charged with electrons, the potential of the charged liquid surface $V_\textrm{e}$ has to be the same everywhere, owing to high mobility of the surface electrons on liquid helium. The value of $V_\textrm{e}$ is determined by voltages applied to the reservoir's bottom and guard electrodes and amount of electrons in the reservoir, and is assumed to be fixed once the device is charged~\cite{remark}. Then, by the definition of capacitance we can write for the total charge $Q$ of electrons in the channel $Q=C_\Sigma(V_\textrm{e}-V_\textrm{b})$. A further simplification can be made by assuming a uniform density distribution of electrons in the channel, that is $Q=-en_sS$, where $n_s$ is the areal density of surface electrons, $e>0$ is the electron charge, and $S$ is the channel area. Such a parallel-plate capacitance approximation is partially justified by a large aspect ratio ($\sim 10$) of a wide and shallow microchannel used in our device. Using $C_\textrm{ch}=\epsilon\epsilon_0 S/d$, where $d=550$~nm is the height of the microchannel in our device, $\epsilon=1.056$ is the dielectric constant of liquid helium, and $\epsilon_0=8.85\times 10^{12}$~F/m is the permittivity of free space, we obtain required relation

\begin{equation}
n_s=\frac{\epsilon\epsilon_0}{\alpha e d} \left( \alpha V_\textrm{ch} + \beta V_\textrm{sg} - V_\textrm{e} \right).
\label{eq:ns}
\end{equation} 

The above equation is very useful to characterize the device and estimate various quantities. For example, the maximum density of electrons corresponds to the condition $V_\textrm{e}=V_\textrm{sg}$, that is electrons remained to be confined across the microchannel by the split-gate potential, from which we find $n_s^{(\textrm{max})}=\epsilon\epsilon_0(V_\textrm{ch}-V_\textrm{sg})/(ed)$. Oppositely, the zero density of electrons in the central microchannel corresponds to the condition $\alpha V_\textrm{ch} + \beta V_\textrm{sg}=V_\textrm{e}$, which determines the threshold value of the channel voltage for a given values of $V_\textrm{sg}$ and $V_\textrm{e}$

\begin{equation}
V_\textrm{ch}^{(\textrm{th})}=\frac{1}{\alpha}V_\textrm{e}- \frac{1-\alpha}{\alpha} V_\textrm{sg}.
\label{eq:thresh}
\end{equation} 
           
\noindent Below this threshold value, the potential at the uncharged surface in the central microchannel $V_\textrm{b}$ is lower than $V_\textrm{e}$, therefore the central microchannel is completely depleted of electrons and the current $I$ in the device is zero. The experimental values of $V_\textrm{ch}^{(\textrm{th})}$ are plotted in Fig.~\ref{fig:2} by a dashed (white) line. By fitting this line using Eq.~(\ref{eq:thresh}) we obtain $V_\textbf{e}=0.92$~V and $\alpha=0.77$ (therefore $\beta=0.23$).

Above the threshold line in the $V_\textrm{sg}$-$V_\textrm{ch}$ plane, see Fig.~\ref{fig:2}, the current in the device is determined by the resistance $R$ of electrons in the microchannel, which in turn depends on the phase of the electron system. For weak confinement of the electron system, which corresponds to lower values of $V_\textrm{ch}$ and more positive values of $V_\textrm{sg}$, the system is in the liquid phase. This corresponds to low resistance $R$ and large current $I$, see Fig.~\ref{fig:2}. For stronger confinement of the electron system, which corresponds to larger values of $V_\textrm{ch}$ and more negative values of $V_\textrm{sg}$, the system undergoes crystallization into WS~\cite{Ikeg2010,ReesPRB2016}. As a result, the resistance $R$ of electrons in the central microchannel increases due to formation of the commensurate dimple lattice, and the measured current $I$ significantly drops. A spectacular behaviour is observed in the intermediate range of voltages, where current $I$ exhibits a fringe pattern, see Fig.~\ref{fig:2}. This phenomenon was identified with the re-entrant melting of WS~\cite{Ikeg2012,Rees2013}. As confining potential, therefore the width of the electron system in the microchannel, is varied by varying voltages applied to the electrodes, the WS in the microchannel undergoes intermittent melting as a result of increased fluctuations of position of electrons between stable configurations corresponding to different number of electron rows across the channel. Therefore, the fringes, which are nearly parallel to the threshold line, see Fig.~\ref{fig:2}, can be identified with different number of electron rows in the microchannel.       

\subsection{Effect of periodic potential}

\begin{figure*}
\includegraphics[width=1.0\textwidth]{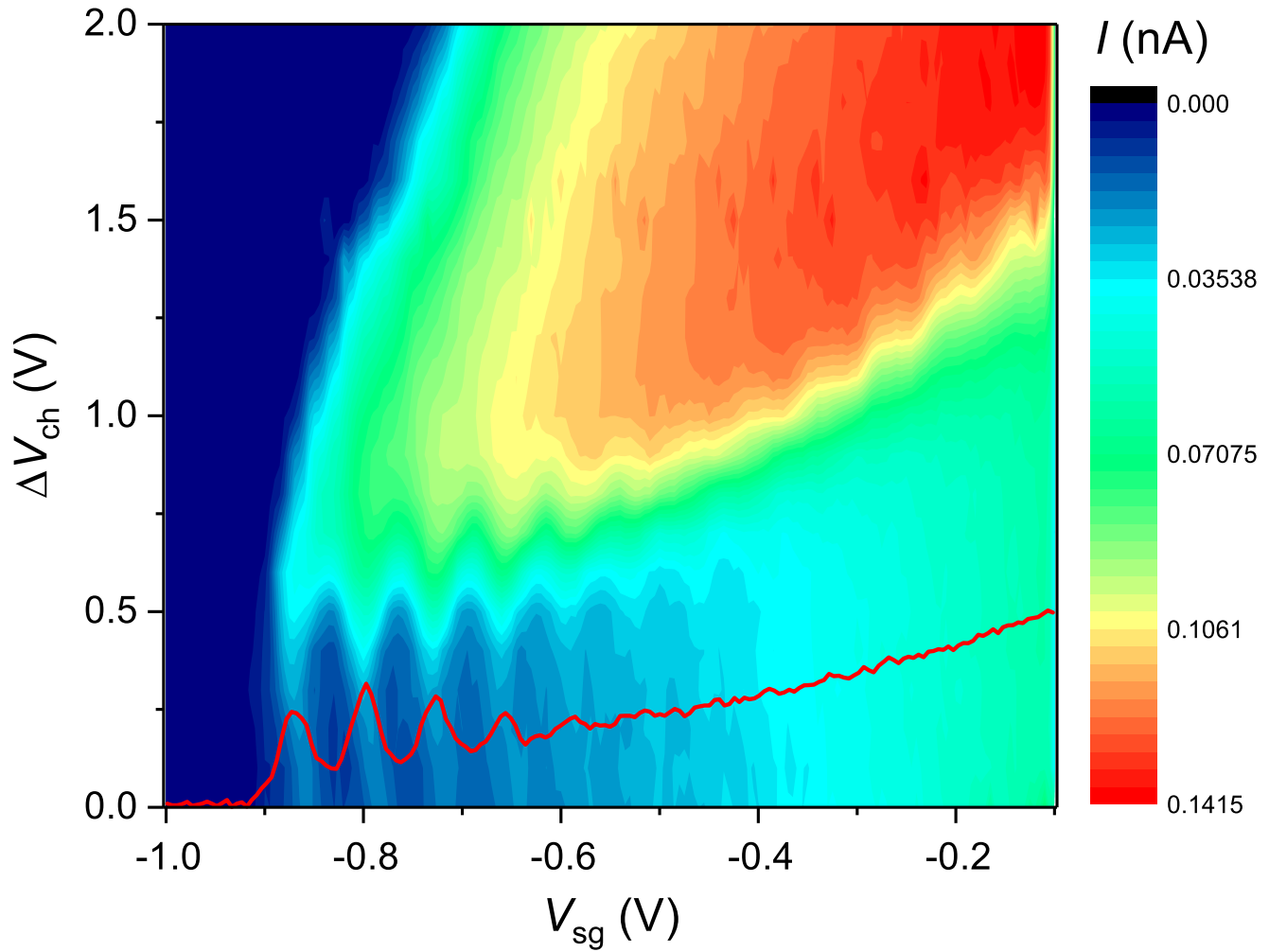}
\caption{(color online) Magnitude of electron current $I$ measured at $T=0.86$~K for driving ac voltage $V_\textrm{ac}=0.5$~mV as a function of split-gate electrode potential $V_\textrm{sg}$ for different values of potential difference $\Delta V_\textrm{ch}=|V_\textrm{ch1}-V_\textrm{ch2}|$ between channel electrodes Ch1 and Ch2. The common bias for two channel electrodes is fixed at $V_\textrm{ch}=1.5$~V. Solid (red) line shows current $I$ versus $V_\textrm{sg}$ measured at a fixed value $\Delta V_\textrm{ch}=0.25$~V.}
\label{fig:3}       
\end{figure*}

Next, we investigate effect of a spatially periodic potential applied to electrons in the microchannel on their current. To do this, we apply potentials $V_\textrm{ch1}=V_\textrm{ch}+\Delta V_\textrm{ch}/2$ and $V_\textrm{ch2}=V_\textrm{ch}-\Delta V_\textrm{ch}/2$ to electrodes Ch1 and Ch2, where $V_\textrm{ch}=1.5$~V is a fixed common bias applied to two electrodes, and $\Delta V_\textrm{ch}$ could be varied from 0 to 2~V. The absolute value of current $I$ in the device is plotted in Fig.~\ref{fig:3} for various values of $\Delta V_\textrm{ch}$ and $V_\textrm{sg}$. For $\Delta V_\textrm{ch}\lesssim 0.7$~V, we observe fringes of current due to the reentrant melting of WS, as described earlier. For the sake of illustration, the solid (red) line plots the measured current $I$ versus $V_\textrm{sg}$ for $\Delta V_\textrm{ch}=0.25$~V. For $\Delta V_\textrm{ch}\gtrsim 0.7$~V, the behaviour becomes drastically different, see Fig.~\ref{fig:3}. The reentrant melting fringes disappear, and measured current $I$ significantly increases to the value comparable to that for electrons in the liquid phase, c.f. Fig.~\ref{fig:2}. This behaviour might suggest that the application of sufficiently strong periodic potential suppresses crystallization of the electron system into the WS phase.

To understand the effect of the spatially periodic potential on the electron system, it is instructive to estimate the variation of the electron density $n_s$ in the central microchannel using the parallel-plate capacitance approximation. As described earlier, the electron density can be estimated as $n_s=\epsilon_0\epsilon \left( V_\textrm{b} - V_\textrm{e} \right)/(\alpha e d)$, where the potential $V_\textrm{b}$ at the uncharged surface of liquid helium in the central microchannel can be calculated numerically using the FEM, see Fig.~\ref{fig:1}(c). We find that at the middle of the channel the density varies nearly sinusoidally with average value $\bar{n}_s$ and amplitude $\Delta n_s$.  In particular, for $V_\textrm{e}=0.92$~V, $V_\textrm{sg}=-0.8$~V, $V_\textrm{ch}=1.5$~V, and $\Delta V_\textrm{ch}=0.7$~V using the above approximation we estimate $\bar{n}_s=3.0\times 10^{13}$~m$^{-2}$ and $\Delta n_s=0.2\times 10^{13}$~m$^{-2}$. For an infinite 2D electron system, the melting of WS is expected to happen when the value of plasma parameter $\Gamma=e^2\sqrt{\pi n_s}/(4\pi\epsilon_0\epsilon k_\textrm{B}T)$ exceeds $130\pm 10$. For $T=0.86$~K, the critical density of electrons corresponds to $n_s=1.4\times 10^{13}$~m$^{-2}$. Therefore, a small variation of electron density due to applied periodic potential estimated above can not cause melting of WS for an infinite electron system. On the other hand, as was pointed out earlier the variation of lateral confinement of the electron system in the microchannel can cause loss of the long-range crystalline order in the quasy-1D WS due to the structural transitions between two stable configurations of the electron lattice corresponding to changing the number $N_y$ of electron rows in the channel by one~\cite{Ikeg2012,Rees2013}. This is exactly the mechanism that explains the phenomenon of re-entrant melting in this system. Therefore, one can expect that variation of $N_y$ along the microchannel caused by the applied periodic potential can induce a similar loss of the long-range positional order, which in turn strongly changes the transport of electron system observed in the experiment. The number of electron rows can be estimated as $N_y=w\sqrt{n_s}$, where $w$ is the width of the electron system in the microchannel. Taking $w$ as the width of the microchannel, the variation of $N_y$ can be estimated as $\Delta N_y=w\left( \sqrt{\bar{n}_s + \Delta n_s} - \sqrt{\bar{n}_s - \Delta n_s} \right)$. Thus, for $V_\textrm{e}=0.92$~V, $V_\textrm{sg}=-0.8$~V, $V_\textrm{ch}=1.5$~V, and $\Delta V_\textrm{ch}=0.7$~V we estimate $\Delta N_y\approx 2$. In other words, the variation of confining potential due to applied periodic potential with $\Delta V_\textrm{ch}=0.7$~V is sufficient to cause the structural transition between $N_y$- and $(N_y+1)$-row configurations, which increases fluctuations in the positions of electrons and suppress nonlinear transport features usually associated with electron system in the WS phase.  

\begin{figure*}
\includegraphics[width=1.0\textwidth]{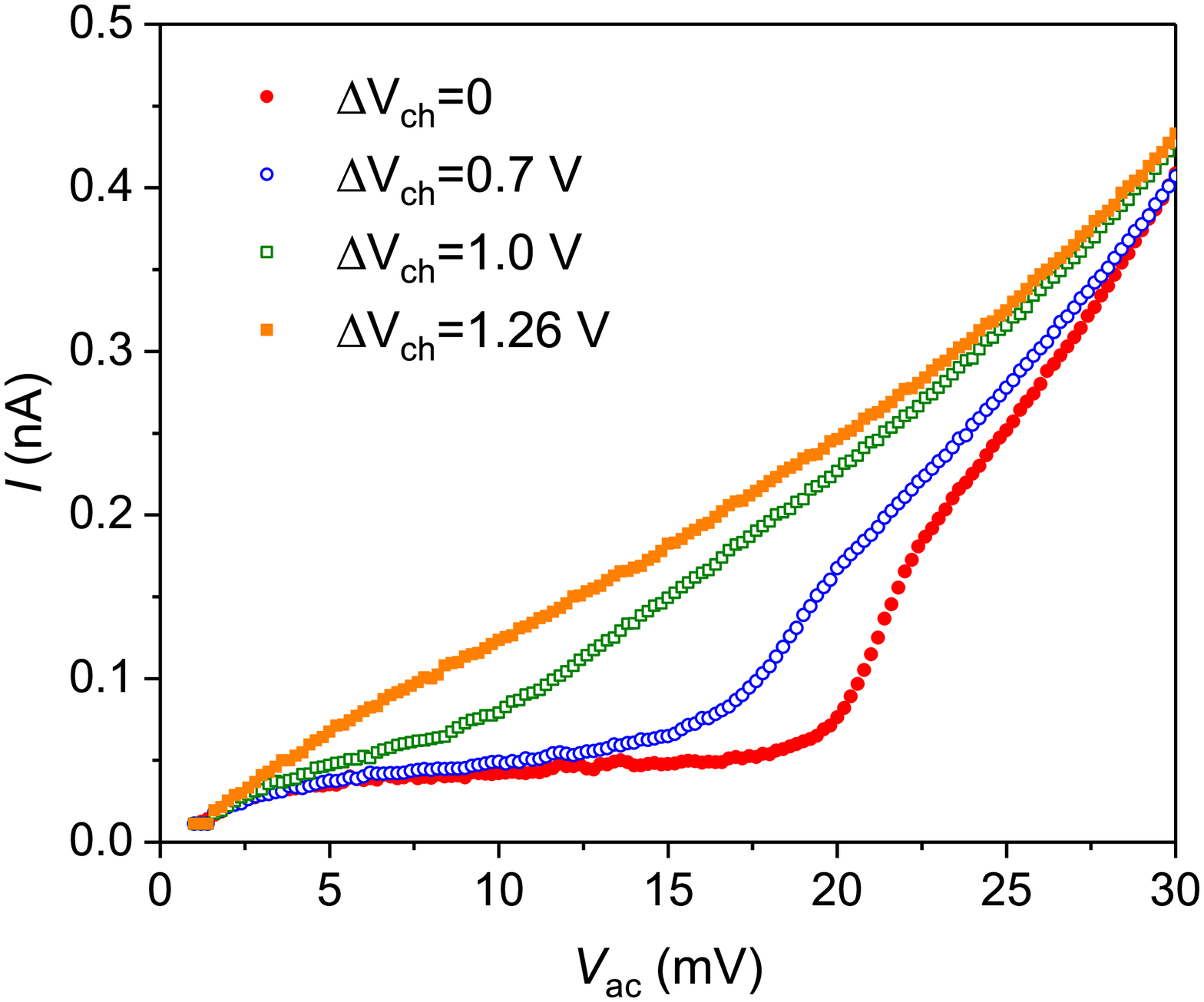}
\caption{(color online) Magnitude of electron current $I$ measured at $T=0.86$~K as a function of peak-to-peak driving ac voltage $V_\textrm{ac}$ at $f=99.5$~kHz for different values of $\Delta V_\textrm{ch}$. The common bias for two channel electrodes was fixed at $V_\textrm{ch}=1.55$~V, while the split-gate voltage was fixed at $V_\textrm{sg}=-0.25$V.}
\label{fig:4}       
\end{figure*}

To confirm suppression of nonlinear transport features associated with the crystalline ordering of the electron system we measured the current $I$ as a function of the driving amplitude $V_\textrm{ac}$ in the presence of periodic potential for different values of $\Delta V_\textrm{ch}$. A typical set of such $IV$-curves is shown in Fig.~\ref{fig:4} for four different values of $\Delta V_\textrm{ch}=0$, 0.7, 1.0, and $1.26$~V. Without periodic potential ($\Delta V_\textrm{ch}=0$), the $IV$-curve clearly shows two characteristic features of nonlinear transport, namely a BC plateau of current due to coherent emission of ripplons by driven WS and a sharp rise of current due to sliding of WS from the commensurate dimple lattice. Application of the periodic potential suppresses both features of nonlinear transport of WS. In particular, for sufficiently large $\Delta V_\textrm{ch}\gtrsim 0.7$~V both features essentially disappear, and the electron transport is close to that of the electron system in liquid phase. This agrees with suppression of re-entrant melting described earlier.                               

\section{Discussion}   

The re-entrant melting, which results from competition between stable configurations corresponding to different numbers of electron rows, is particularly important in studies of finite-size crystalline systems where the spatial order of particles is strongly affected by their confinement~\cite{ReesPRB2016}. The interplay between the electron lattice configuration and confining potential is an interesting problem of structural phase transitions~\cite{FishmanPRB2008}. In our experiment, the confinement is spatially modulated by the external periodic potential of varying strength. A characteristic feature of our observation is a certain (threshold) value of the amplitude of periodic potential above which the nonlinear transport of the electron system associated with its crystalline ordering is suppressed. Our estimations presented above show that this corresponds to about 10$\%$-variation of the electron density in microchannel and variation of the number $N_y$ of electron rows across the channel of the order one. More accurate estimations could be done by calculating the distribution of electrical potential and electron density across the microchannel ($y$-direction) using the FEM~\cite{Rees2013,Ikeg2015}. However, to take a proper account for the granular nature of electrons the molecular dynamics (MD) calculation are preferable~\cite{ReesPRB2016}. Therefore, we did not try to improve the continuous density approximation model used in the previous section. The MD calculations for electron system in our device is currently under development.

As demonstrated in our work, the employed microchannel device can be used to study structural phase transitions in a quasy-1D electron systems. Also, we are interested to use similar devices to study the FK model employing a 1D chain of electrons subject to the periodic potential. Of particular interest is to realize an incommensurate case when the ratio of the mean distance between electrons to the spatial period of potential is equal to the "golden ratio", $(\sqrt{5}+1)/2$. This is subject of our future experimental efforts. 

\section{Summary}

We have investigated the transport properties of a quasi-1D WS on the surface of liquid helium confined in a long 5~$\mu$m-wide microchannel and subjected to an applied electrostatic potential with periodicity of 1~$\mu$m along the channel. The nonlinear features of WS transport were found to be suppressed by increasing the potential amplitude. We attribute this observation to structural transitions and suppression of the crystalline ordering of the electron system induced by the spatially modulated confinement.

%

\begin{acknowledgements}
The work was supported by an internal grant from the Okinawa Institute of Science and Technology (OIST) Graduate University. A. O. B. was partially supported by JSPS KAKENHI Grant Number JP18K13506.

\end{acknowledgements}



\begin{thebibliography}{}

\bibitem{Van2013} A. Vanossi, N. Manini, M. Urbakh, S. Zapperi, and E. Tosatti, Rev. Mod. Phys. \textbf{85}, 529 (2013).

\bibitem{FK-book} O. M. Braun and Y. S. Kivshar, {\it The Frenkel-Kontorova Model: Concepts, Methods, and Applications} (Springer-Verlag, Berlin, 2004).

\bibitem{Die2004} Dienwiebel, M., G. S. Verhoeven, N. Pradeep, J.W. M. Frenken, J. A. Heimberg, and H.W. Zandbergen, Phys. Rev. Lett. 92, 126101, 2004.

\bibitem{Soc2006} Socoliuc, A., E. Gnecco, S. Maier, O. Pfeiffer, A. Baratoff, R. Bennewitz, and E. Meyer, Science 313, 207 (2006).

\bibitem{Bil2015} A. Bylinskii, D. Gangloff, and V. Vuletic, Science \textbf{349}, 1115 (2015). 

\bibitem{Bil2016} A. Bylinskii, D. Gangloff, I. Counts, and V. Vuletic, Nature Materials \textbf{15}, 717 (2016). 

\bibitem{Kri1996} A. Kristensen, K. Djerfi, P. Fozooni, M. J. Lea, P. J. Richardson, A. Santrich-Badal, A. Blackburn, and R.W. van der Heijden, Phys. Rev. Lett. 77, 1350 (1996).

\bibitem{Dyk1997} M. I. Dykman and Y. G. Rubo, Phys. Rev. Lett. 78, 4813 (1997).

\bibitem{Shir1995} K. Shirahama and K. Kono, Phys. Rev. Lett. 74, 781 (1995).

\bibitem{Andrei} E. Y. Andrei, {\it Electrons on Helium and Other Cryogenic Substrates} (Kluwer Academic, Dordrecht, 1997).



\bibitem{VanossiPRL2008} A. Vanossi, N. Manini, F. Caruso, G. E. Santoro, and E. Tosatti, Phys. Rev. Lett. \textbf{99}, 206101 (2007). 

\bibitem{WangAPL2008} C.-L. Wang, W.-S. Duan, X.-R. Hong, and J.-M. Chen, Appl. Phys. Lett. \textbf{93}, 153116 (2008). 

\bibitem{Brad2011} F. R. Bradbury, M. Takita, T. M. Gurrieri, K. J. Wilkel, K. Eng, M. S. Carroll, and S. A. Lyon, Phys. Rev. Lett. 107, 266803 (2011).

\bibitem{Rees2011} D. G. Rees, I. Kuroda, C. A. Marrache-Kikuchi, M. Hofer, P. Leiderer, and K. Kono, Phys. Rev. Lett 106, 026803 (2011).

\bibitem{Rees2012} D. G. Rees, I. Kuroda, C. A. Marrache-Kikuchi, M. Hofer, P. Leiderer, and K. Kono, L. Low Temp. Phys. 166, 107 (2012).

\bibitem{Ikeg2010} H. Ikegami, H. Akimoto, and K. Kono, Phys. Rev. B \textbf{82}, 201104(R) (2010).

\bibitem{Ikeg2012} H. Ikegami, H. Akimoto, D. G. Rees, and K. Kono, Phys. Rev. Lett. 109, 236802 (2012).

\bibitem{Rees2013} D. G. Rees, H. Ikegami, and K. Kono, J. Phys. Soc. Jpn. 82, 124602 (2013).

\bibitem{Rees2016} D. G. Rees, N. R. Beysengulov, J. J. Lin, and K. Kono, Phys. Rev. Lett. 116, 206801 (2016).

\bibitem{BadrPRB2016} A. O. Badrutdinov, A. V. Smorodin, D. G. Rees, J.-Y. Lin and D. Konstantinov, Phys. Rev. B \textbf{94}, 195311 (2016).

\bibitem{Ikeg2015} H. Ikegami, H. Akimoto, and K. Kono, J. Low Temp. Phys. \textbf{179}, 251 (2015).

\bibitem{remark} Occasionally, loss of electrons from the device is observed, which is reflected in discontinuous jumps of the measured current $I$. Such data are not considered here.

\bibitem{ReesPRB2016} D. G. Rees, N. R. Beysenulov, Y. Teranishi, C.-S. Tsao, S.-S. Yeh, S.-P. Chiu, Y.-H. Lin, D. A. Tayurskii, J.-J. Lin, and K. Kono, Phys. Rev. B \textbf{94}, 045139 (2016).

\bibitem{FishmanPRB2008} S. Fishman, G. De Chiara, T. Calarco, and G. Morigi, Phys. Rev. B \textbf{77}, 064111 (2008).




\end{thebibliography}


\end{document}